\documentclass[aps,prd,twocolumn]{revtex4-1}

\usepackage{eurosym}
\usepackage{amssymb}
\usepackage{amsfonts}
\usepackage{amsmath}
\usepackage{graphicx}
\usepackage{bm}
\usepackage{color}
\usepackage[usenames,dvipsnames]{xcolor}
\usepackage[dvips]{epsfig}
\usepackage[dvips]{graphicx}
\usepackage{float}
\usepackage{epstopdf}
\usepackage[colorlinks=true,pdfstartview=FitV,linkcolor=blue,citecolor=blue,urlcolor=blue,breaklinks=true]{hyperref}
\usepackage{wasysym}
\usepackage{blindtext}

\begin{document}

\title{An exact Schwarzschild-like solution in a bumblebee gravity model}
\author{R. Casana}
\email{rodolfo.casana@gmail.com}
\author{A. Cavalcante}
\email{andre$_$cavs@hotmail.com}
\affiliation{Departamento de F\'{\i}sica, Universidade Federal do Maranh\~{a}o,
65080-805, S\~{a}o Lu\'{\i}s, Maranh\~{a}o, Brazil.}
\author{F. P. Poulis}
\email{fppoulis@gmail.com}
\affiliation{Coordena\c{c}\~{a}o do Curso Interdisciplinar em Ci\^{e}ncia e Tecnologia, Universidade
Federal do Maranh\~{a}o, 65080-805, S\~{a}o Lu\'{\i}s, Maranh\~{a}o, Brazil.}
\author{E. B. Santos}
\email{esdras.bsantos@hotmail.com}
\affiliation{Departamento de F\'{\i}sica, Universidade Federal de Pernambuco, 50670-901,
Recife, Pernambuco, Brazil.}

\begin{abstract}
{We have obtained an exact vacuum solution from a gravity sector contained in the minimal standard-model extension.} The theoretical
model assumes a Riemann spacetime coupled to the bumblebee field which is
responsible for the spontaneous Lorentz symmetry breaking. {The solution achieved in} a static and spherically symmetric scenario establishes a Schwarzschild-like black hole. In order to study the effects of the
spontaneous Lorentz symmetry breaking, we have investigated some classics
tests including the advance of the perihelion, bending of light and
Shapiro's time-delay. Furthermore, we have computed some upper-bounds from
which the most stringent one attains a sensitivity at the  $10^{-13}$ level.
\end{abstract}

\maketitle



\section{Introduction\label{Intro}}

General Relativity (GR) and Standard Model (SM) of particle physics {%
{}are examples of successfully field theories describing
nature.} The former describes gravitation at a classical level, and the
latter describes particles and the other three fundamental interactions at a
quantum level. The unification between these two theories is a fundamental
seeking, and this achievement will conduct us necessarily to a deeper
understanding of nature.

In the pursuit of this unification some theories of
quantum gravity (QG) have already been proposed, but direct tests of their
properties are currently beyond the energy scale of current
experiments because they would be observed at the Planck
scale ($\sim 10^{19}$ GeV). However, it is possible
that some signals of QG can emerge at sufficiently low energy scales and its
effects could be observed in experiments carried out at current energy
scales. One of these signals would be associated with
the breaking of Lorentz symmetry.

Lorentz symmetry {{}breaking in nature} is an interesting
idea to be considered because it arises {{}as a possibility
in the context of string theory \cite{PRD39.1989,PRD55.1997}, noncommutative
field theories \cite{PRL87.2001} or loop quantum gravity theory \cite{LoopQG}%
}. This fact suggests the seeking {{}by evidences of
Lorentz violation} is a promising way to investigate signals of the
existence of an underlying theory of quantum gravity at the Planck scale.

The standard model extension (SME) is an effective field {{}%
theory describing at low} energies the general relativity and {%
{}the standard model,} including in its structures
additional terms containing information about the Lorentz violation {%
{}occurring at the} Planck scale \cite{PRD69.2004}. {%
{}The electromagnetic sector of the SME has been
extensively studied in literature \cite{PLB511.2001,EPJP129.2014,JMP40.1999,
PRD86.2012,PRD68.2003.2004,JMP.2007,KM,CFJ,n11,n12,n13,PLB628.2005}; the electroweak sector in Refs. \cite{EWeak}, some aspects of the strong sector \cite{Strong} and the hadronic physics \cite{Hadron} as well. Furthermore, some effects of Lorentz violation in the gravitational sector have been studied in \cite{n14,n16,n17,n18,PRD74.2006}, specifically the case of the
gravitational waves were analyzed in Ref. \cite{Gwaves}}.

{The aim of the manuscript is to obtain a spherically
symmetric exact solution of the Einstein equations in the presence of a
spontaneous breaking of Lorentz symmetry because of the nonzero vacuum
expectation value of the bumblebee field and its influence in some
well-known experimental tests of general relativity: The advance of
perihelion of inner planets, the bending of light and time delay effect owing to
curvature. All tests we analyzed allow to estimate some upper-bounds for the
Lorentz-violating parameter involved. We will adopt the metric signature $%
(-+++)$ and also all quantities involved are expressed in natural units }$%
\left( \hslash =c=1\right) ${{}. When necessary the
physical constants will be written explicitly. The manuscript is developed
in the following manner: In Sec. \ref{sec2}, it is presented a general
geometrical framework allowing the existence of nonzero vacuum expectation
values promoting the spontaneous breaking of local Lorentz invariance.
Furthermore, we have computed the modified Einstein's equation generated by
the bumblebee gravity. In Sec. \ref{sec3}, we solve the modified Einstein's
equation seeking for spherically symmetric solutions. In Sec. \ref{CT}, we
study the effects of Lorentz violation in some classical tests of general
relativity and we use experimental data {to} establish some upper-bounds on
the Lorentz-violating parameter involved. Finally, in Sec. \ref{end}, we
give our conclusions and remarks}

\section{The Theoretical Framework \label{sec2}}

{The focus of this work is to study spherically symmetric
vacuum solutions in the context of an extended gravitational model including
Lorentz-violating terms. Consequently, we study the effects of Lorentz
violation in some classical tests of general relativity.} For this purpose,
we consider the bumblebee model, which is a known example of a gravity model
that extends the standard formalism of GR, {{}where under a
suitable potential the bumblebee vector field $B_{\mu }$ acquires a
nonvanishing vacuum expectation value (VEV) inducing a spontaneous Lorentz
symmetry breaking (LSB).}

In order to investigate the spontaneous Lorentz symmetry breaking in the
extended gravitational sector, we consider the special class of theories in
which the Lorentz violation arises from the dynamics of a single vector
{$B_{\mu}$} that acquires a nonzero vacuum expectation value. These theories
are called bumblebee models and are among the simplest examples of field
theories with spontaneous Lorentz and diffeomorphism violations. It is
well-known in the literature that the local LSB is always accompanied by
diffeomorphism violation \cite{PRD71.2005}. In this
scenario, the spontaneous LSB is triggered by a potential whose functional
form possesses a minimum which ensures the breaking of the $U(1)$ symmetry. In
general, the action for a single bumblebee field $B_{\mu }$ coupled to
gravity and matter can be written as
\begin{eqnarray}
S_{B} &=&\int d^{4}x~\mathcal{L}_{B}  \label{4} \\
&=&\int d^{4}x\left( \mathcal{L}_{g}+\mathcal{L}_{gB}+\mathcal{L}_{\text{K}}+%
\mathcal{L}_{V}+\mathcal{L}_{\text{M}}\right) .  \notag
\end{eqnarray}%
In Riemann spacetime, $\mathcal{L}_{g}$ is the pure gravitational
Einstein-Hilbert term which may also include the cosmological constant, $%
\mathcal{L}_{gB}$ describes the gravity-bumblebee coupling, $\mathcal{L}_{%
\text{K}}$ contains the bumblebee kinetic and any self-interaction terms, $%
\mathcal{L}_{V}$ correspond to the potential, which includes terms that
trigger the spontaneous Lorentz violation, {{}and $\mathcal{%
L}_{\text{M}}$ defines the matter and others field contents and their
couplings to the bumblebee field. By considering the case of a spacetime
with null torsion and null cosmological constant ($\Lambda =0$), we
introduce the following Lagrangian density }
\begin{eqnarray}
\mathcal{L}_{B} &=&\frac{e}{2\kappa }R+\frac{e}{2\kappa }\xi {B^{\mu }B^{\nu
}}R_{\mu \nu }-\frac{1}{4}eB_{\mu \nu }B^{\mu \nu }  \notag \\[0.08in]
&&-eV(B^{\mu}) +\mathcal{L}_{\text{M}}\text{,}  \label{6}
\end{eqnarray}%
being $e\equiv \sqrt{-g}$ the determinant of the vierbein and $\xi $ the
real coupling constant {{}(with mass dimension }$-1$) which
controls the nonminimal gravity-bumblebee interaction. The corresponding
bumblebee field strength is defined as
\begin{equation}
B_{\mu \nu }=\partial _{\mu }B_{\nu }-\partial _{\mu }B_{\nu }\text{,}
\end{equation}%
where $B_{\mu }$ has {{}mass dimensions }$1$. We point out
that some bumblebee models involving nonzero torsion in more general context
are investigated in Refs. \cite{PRD69.2004,PRD71.2005}.

For our purposes, the particular form of the potential $V(B_{\mu })$ in Eq. (%
\ref{6}) driving its dynamics is irrelevant, but it is important to
emphasize that it must be formed from scalar combinations of the bumblebee
field $B_{\mu }$ and the metric $g_{\mu \nu }$. {{}In any
case, we choose a potential $V$ providing a nonvanishing VEV for $B_{\mu }$
which could have the following general functional form}
\begin{equation}
V\equiv V(B^{\mu }B_{\mu }\pm b^{2}),  \label{V}
\end{equation}%
{{}where $b^{2}$ is a positive real constant. Some
qualitative features of the symmetry breaking potential have been explored}
in Refs. \cite{PRD39.1989,PRD69.2004,PRD71.2005,GRG37.2005}. It follows the
VEV of the bumblebee field is determined when $V(B^{\mu }B_{\mu }\pm
b^{2})=0 $ implying that the condition
\begin{equation}
B^{\mu }B_{\mu }\pm b^{2}=0\text{,}  \label{vacuum}
\end{equation}%
must be satisfied. {{}This is solved when the field $B^{\mu
}$ acquires a non null vacuum expectation value }
\begin{equation}
\left\langle B^{\mu }\right\rangle =b^{\mu }\text{,}  \label{vev}
\end{equation}%
{{}where the vector }$b^{\mu }${{}\ is a
function of the spacetime coordinates such that }$b^{\mu }b_{\mu }=\mp
b^{2}=\text{const.}${{}, then the nonnull vector background} $b^{\mu
} $\ spontaneously breaks the Lorentz symmetry. We note the $\pm $ signs in
the potential (\ref{V}) determine {{}whether the field }$%
b^{\mu }${{}\ is timelike or spacelike.}

On the other hand, the Lorentz-violating contribution to the gravitational
sector provided by the minimal SME is
\begin{equation}
S_{LV}=\frac{1}{2\kappa }\int d^{4}x\sqrt{-g}(uR+s^{\mu \nu }R_{\mu \nu
}+t^{\mu \nu \alpha \beta }R_{\mu \nu \alpha \beta })\text{,}  \label{3}
\end{equation}%
where $u$, $s^{\mu \nu }$ and $t^{\mu \nu \alpha \beta }$ are real and
dimensionless tensors carrying on information about Lorentz violation. It is
possible to {establish} a correspondence between the bumblebee action (\ref%
{4}) and the Lorentz-violating action (\ref{3}) {by considering} the
following correspondence of the underlying bumblebee field and the metric
with the Lorentz-violating tensors $u$, $s^{\mu \nu }$ and $t^{\mu \nu
\alpha \beta }$:
\begin{eqnarray}
u=\frac{1}{4}\xi B^{\mu }B_{\mu }\,,\;\, s^{\mu \nu }=\xi \left( B^{\mu
}B^{\nu }-\frac{1}{4}g^{\mu \nu }B^{\alpha }B_{\alpha }\right) ,\quad
\label{rel}
\end{eqnarray}
\begin{eqnarray}
t^{\mu \nu \alpha \beta }&=& 0\text{,}  \label{rel0}
\end{eqnarray}%
{{}with $s^{\mu \nu }$ being traceless \cite%
{PRD69.2004,PRD71.2005,PRD74.2006,PRD90.2014}.}

{The next step is to establish the fields equations from
the action (\ref{4}) aiming at finding vacuum solutions in the context of
extended gravitational sector.}

\subsection{The Field Equations}

The Lagrangian density (\ref{6}) yields the extended Einstein equations
\begin{equation}
G_{\mu \nu }=R_{\mu \nu }-\frac{1}{2}Rg_{\mu \nu }=\kappa T_{\mu \nu }\text{,%
}  \label{FE}
\end{equation}%
where $G_{\mu \nu }$ is the Einstein tensor and $T_{\mu \nu }$ {%
{}is the total energy-momentum tensor arising from the
matter sector (}$T_{\mu \nu }^{\text{M}}${{}) and the
contributions of the bumblebee field ($T_{\mu \nu }^{B}$), so we write}%
\begin{equation}
T_{\mu \nu }=T_{\mu \nu }^{\text{M}}+T_{\mu \nu }^{B}
\end{equation}%
{{}with}
\begin{eqnarray}
T_{\mu \nu }^{B} &=&-B_{\mu \alpha }B_{~\nu }^{\alpha }-\frac{1}{4}B_{\alpha
\beta }B^{\alpha \beta }g_{\mu \nu }-Vg_{\mu \nu }+2V^{\prime }B_{\mu
}B_{\nu }  \notag \\[0.08in]
&&+\frac{\xi }{\kappa }\left[ \frac{1}{2}B^{\alpha }B^{\beta }R_{\alpha
\beta }g_{\mu \nu }-B_{\mu }B^{\alpha }R_{\alpha \nu }-B_{\nu }B^{\alpha
}R_{\alpha \mu }\right.  \notag \\[0.08in]
&&+\frac{1}{2}\nabla _{\alpha }\nabla _{\mu }\left( B^{\alpha }B_{\nu
}\right) +\frac{1}{2}\nabla _{\alpha }\nabla _{\nu }\left( B^{\alpha }B_{\mu
}\right)  \notag \\[0.08in]
&&\left. -\frac{1}{2}\nabla ^{2}\left( B_{\mu }B_{\nu }\right) -\frac{1}{2}%
g_{\mu \nu }\nabla _{\alpha }\nabla _{\beta }\left( B^{\alpha }B^{\beta
}\right) \right] \text{.}
\end{eqnarray}%
The prime denotes differentiation with respect to the argument, as usual. {%
{}Similarly, Eq. (\ref{6}) provides the following equation
of motion for the bumblebee field,}
\begin{equation}
\nabla ^{\mu }B_{\mu \nu }=J_{\nu }\text{,}  \label{curr}
\end{equation}%
where $J_{\nu }=J_{\nu}^{B}+J_{\nu }^{\text{M}}$, with $J_{\nu }^{\text{M}}$
being associated with the matter sector (acting as a source of the bumblebee
field) and $J_{\nu }^{B}$ {{}a partial current that arises
from the bumblebee self-interaction} given explicitly as
\begin{equation}
J_{\nu }^{B}=2V^{\prime }B_{\nu }-\frac{\xi }{\kappa }B^{\mu }R_{\mu \nu }%
\text{.}
\end{equation}

{Taking the covariant divergence} on the {extended}
Einstein equations (\ref{FE}) and using the {{}contracted}
Bianchi identities ($\nabla ^{\mu }G_{\mu \nu }=0$) {{},
this leads to condition}
\begin{equation}
\nabla ^{\mu }T_{\mu \nu }=0\text{,}
\end{equation}%
which gives the covariant conservation law for the {{}total}
energy-momentum tensor $T_{\mu \nu }$.

{The trace of Eq. (\ref{FE}) reads}
\begin{eqnarray}
R &=&-\kappa T^{\text{M}}+4\kappa V-2\kappa V^{\prime }B_{\mu }B^{\mu }
\notag \\[0.08in]
&&+\xi \left[ \frac{1}{2}\nabla ^{2}\left( B_{\mu }B^{\mu }\right) +\nabla
_{\alpha }\nabla _{\beta }\left( B^{\alpha }B^{\beta }\right) \right] \text{,%
}  \label{rrr}
\end{eqnarray}%
where $T^{\text{M}}\equiv g^{\mu \nu }T_{\mu \nu }^{\text{M}}$, {and {%
{}substituting it in Eq. (\ref{FE}) we obtain} the
trace-reversed version}
\begin{eqnarray}
R_{\mu \nu } &=&\kappa \left( T_{\mu \nu }^{\text{M}}-\frac{1}{2}g_{\mu \nu
}T^{\text{M}}\right) +\kappa T_{\mu \nu }^{B}+2\kappa g_{\mu \nu }V  \notag
\\[0.08in]
&&-\kappa B_{\alpha }B^{\alpha }g_{\mu \nu }V^{\prime }+\frac{\xi }{4}g_{\mu
\nu }\nabla ^{2}\left( B_{\alpha }B^{\alpha }\right)  \notag \\[0.08in]
&&+\frac{\xi }{2}g_{\mu \nu }\nabla _{\alpha }\nabla _{\beta }(B^{\alpha
}B^{\beta })\text{.}  \label{ricci}
\end{eqnarray}%
Note that {if both} the bumblebee field $B_{\mu }$ and the potential $%
V(B_{\mu })$ vanishes, Eq. (\ref{ricci}) recovers the usual GR equations, as
expected.

\section{A spherically symmetric solution in the LSB scenario \label{sec3}}

We focus on a vacuum solution, i.e., the one describing an empty space
surrounding a gravitating body by imposing $T_{\mu \nu}^{\text{M}}=0$. We
note the potential in (\ref{6}) vanishes when Eq. (\ref{vev}) is satisfied,
which characterizes the vacuum. We will assume this scenario in the
remainder of this manuscript.

{{}Specifically, we are interested in vacuum solutions
induced by the LSB when the bumblebee field }$B_{\mu }${{}
remains frozen in its vacuum expectation value }$b_{\mu }${%
{}. Similar hypothesis was used in} Ref. \cite{PRD72.2005}.
{In this way, the bumblebee field is fixed to be }%
\begin{equation}
B_{\mu }=b_{\mu }\text{,}  \label{rxd}
\end{equation}%
{{}consequently, we have }$V=0${{}\ and }$%
V^{\prime }=0${{}, which when substituted in Eq. (\ref%
{ricci}) provides the extended Einstein equations in vacuum }%
\begin{eqnarray}
0 &=&\bar{R}_{\mu \nu }  \label{R1} \\[0.08in]
&=&R_{\mu \nu }+\kappa b_{\mu \alpha }b_{~\,\nu }^{\alpha }+\frac{\kappa }{4}%
b_{\alpha \beta }b^{\alpha \beta }g_{\mu \nu }+\xi b_{\mu }b^{\alpha
}R_{\alpha \nu }  \notag \\
&&+\xi b_{\nu }b^{\alpha }R_{\alpha \mu }-\frac{\xi }{2}b^{\alpha }b^{\beta
}R_{\alpha \beta }g_{\mu \nu }-\frac{\xi }{2}\nabla _{\alpha }\nabla _{\mu
}\left( b^{\alpha }b_{\nu }\right)  \notag \\
&&-\frac{\xi }{2}\nabla _{\alpha }\nabla _{\nu }\left( b^{\alpha }b_{\mu
}\right) +\frac{\xi }{2}\nabla ^{2}\left( b_{\mu }b_{\nu }\right) \text{,}
\notag
\end{eqnarray}%
{where $b_{\mu \nu }\equiv \partial _{\mu }b_{\nu
}-\partial _{\nu }b_{\mu }$ is the field strength of the vector $b_{\nu}$.}

In order to obtain a static, spherically symmetric vacuum solution to the
extended Einstein equations, we assume a spacetime driven by a Birkhoff
metric $g_{\mu \nu }=$\thinspace diag$\left( -e^{2\gamma },e^{2\rho
},r^{2},r^{2}\sin ^{2}\theta \right) $, with $\gamma $ and $\rho $ being
functions of $r$. {{}Hereafter, we consider a spacelike
background }$b_{\mu }${{}\ assuming the form}
\begin{equation}
b_{\mu }=\left( 0,b_{r}(r) ,0,0\right) \text{.}  \label{br}
\end{equation}%
{{}Moreover, once we have assumed a background field in the
form (\ref{br}), it follows that all components of the corresponding field
strength vanishes, i.e., }$b_{\mu \nu }=0${{}.}

{Now by using the condition }$b^{\mu }b_{\mu }=b^{2}={\mbox{const.}}${%
, we determine the explicit form of {the radial} background
field}
\begin{equation}
b_{r}(r) =\left\vert b\right\vert e^{\rho }\text{,}  \label{bb}
\end{equation}%
{It is easy to verify {that the} background given by Eq. (\ref{bb}) is not covariantly
constant, i.e., we have some nonvanishing values for }$\nabla _{\mu }b_{\nu
} ${{}. \ It is worthwhile to point the difference with the
{proposal} analyzed in Ref. \cite{PRD72.2005} which {assumes} the condition }$%
\nabla _{\mu }b_{\nu }=0${{}.}

{Next, we proceed} to solve for the functions $\gamma (r) $ and $\rho (r) $.
For this, we take {the extended Einstein equations in vacuum with our metric
ansatz to get} the following nonvanishing components for the tensor (\ref{R1}%
):
\begin{eqnarray}
\bar{R}_{tt} &=&\left( 1+\frac{\ell }{2}\right) R_{tt}+\frac{\ell }{r}\left(
\partial _{r}\gamma +\partial _{r}\rho \right) e^{2(\gamma -\rho )}\text{,}%
\quad \quad  \label{r1} \\[0.08in]
\bar{R}_{rr} &=&\left( 1+\frac{3\ell }{2}\right) R_{rr}\text{,}  \label{r2}
\\[0.06in]
\bar{R}_{\theta \theta } &=&\left( 1+\ell \right) R_{\theta \theta }-\ell
\left( \frac{1}{2}r^{2}e^{-2\rho }R_{rr}+1\right) \text{,}  \label{r3} \\%
[0.08in]
\bar{R}_{\phi \phi } &=&\sin ^{2}\theta \bar{R}_{\theta \theta }\text{,}
\label{r4}
\end{eqnarray}%
where we have defined the constant $\ell =\xi b^{2}$. The components of the
Ricci tensor $R_{\mu \nu }$ {that appear above are given by}%
\begin{eqnarray}
R_{tt} &=&e^{2(\gamma -\rho )}\left[ \partial _{r}^{2}\gamma +\left(
\partial _{r}\gamma \right) ^{2}-\partial _{r}\gamma \partial _{r}\rho +%
\frac{2}{r}\partial _{r}\gamma \right] \text{,}\quad \quad \\[0.06in]
R_{rr} &=&-\partial _{r}^{2}\gamma -\left( \partial _{r}\gamma \right)
^{2}+\partial _{r}\gamma \partial _{r}\rho +\frac{2}{r}\partial _{r}\rho
\text{,} \\[0.08in]
R_{\theta \theta } &=&e^{-2\rho }\left[ r\left( \partial _{r}\rho -\partial
_{r}\gamma \right) -1\right] +1\text{.}
\end{eqnarray}%
Note that, {as in a scenario without a background field, $\bar{R}_{\mu \nu }$
also has only three diagonal independent components}.

According to Eq. (\ref{R1}), {{}each of these components
{vanishes} independently so we do the following combination:}
\begin{equation}
r^{2}e^{-2\gamma }\bar{R}_{tt}-r^{2}e^{-2\rho }\bar{R}_{rr}+2\bar{R}_{\theta
\theta }=0\text{,}
\end{equation}%
{{}which yields a differential equation for the function $%
\rho(r)$,}
\begin{equation}
\partial _{r}\left( re^{-2\rho }\right) \left( 1+\ell \right) =1\text{.}
\end{equation}%
{{}It is easy to show the solution is}
\begin{equation}
e^{2\rho }=\left( 1+\ell \right) \left( 1-\frac{\rho _{0}}{r}\right) ^{-1}%
\text{,}  \label{s2}
\end{equation}%
{where for now $\rho _{0}$ is some} arbitrary constant.

{With the aim to find the function $\gamma (r) $ we
consider the following combination}
\begin{equation}
r^{2}e^{-2\gamma }\bar{R}_{tt}-\left( 1+\frac{2}{\ell }\right) \bar{R}%
_{\theta \theta }=0\text{,}
\end{equation}%
which provides%
\begin{eqnarray}
0&=&\left( 2+3\ell \right) \left( 1+\ell \right) r\partial _{r}\gamma
+\left( 1+\ell \right) \left( 2+\ell \right)  \notag \\[0.15cm]
&&+\left( \ell ^{2}-\ell -2\right) r\partial _{r}\rho -\left( 2+\ell \right)
e^{2\rho } \text{.}
\end{eqnarray}%
{By substituting Eq. (\ref{s2}), we obtain an explicit
differential equation for $\gamma (r) $,}
\begin{equation}
\left( 2+3\ell \right) \left[ \left( \rho _{0}-r\right) \partial _{r}\gamma +%
\frac{\rho _{0}}{2r}\right] =0\text{,}
\end{equation}%
whose solution, written in a convenient form, is given by%
\begin{equation}
e^{2\gamma }=e^{-2\gamma _{0}}\left( 1-\frac{\rho _{0}}{r}\right) \text{,}
\label{s1}
\end{equation}%
where $e^{-2\gamma _{0}}$ {{}is a constant which can be
removed by means of the rescaling} $t\rightarrow e^{\gamma_{0}}t$. {%
{}It {can be} verified in fact the solutions } (\ref{s2}) and (%
\ref{s1}) actually satisfy the set of Eqs. (\ref{r1})-(\ref{r4}).

{Finally, we write down the LSB spherically symmetric
solution}
\begin{eqnarray}
ds^{2} &=&-\left( 1-\frac{2M}{r}\right) dt^{2}+\left( 1+\ell \right) \left(
1-\frac{2M}{r}\right) ^{-1}dr^{2}  \notag \\[0.08in]
&&+r^{2}d\theta ^{2}+r^{2}\sin ^{2}\theta d\phi ^{2}\text{,}  \label{line}
\end{eqnarray}%
{{}where we have conveniently identified $\rho _{0}\equiv
2M $ ($M=G_{N}m$ is the usual geometrical mass) such that in the limit $\ell
\rightarrow 0~(b^{2}\rightarrow 0)$ we recover the usual Schwarzschild
metric. The metric} (\ref{line}) represents a purely radial LSB solution
outside a spherical body {{}characterizing a modified black
hole solution. Furthermore, we compute the Kretschmann scalar}
\begin{equation}
R_{\mu \nu \alpha \beta }R^{\mu \nu \alpha \beta }=\frac{4\left(
12M^{2}+4\ell Mr+\ell ^{2}r^{2}\right) }{r^{6}\left( \ell +1\right) ^{2}}%
\text{,}  \label{inv}
\end{equation}%
{{}which clearly differs {from} the Schwarzschild Kretschmann
invariant for nonnull $\ell $. This ensures the metric (\ref{line}) {is a} true
solution containing Lorentz-violating corrections, i.e., it means there
exists no coordinate transformation connecting the metric (\ref{line}) to
the Schwarzschild one, otherwise, the scalar invariant (\ref{inv}) would be
the same for both metrics. We observe, for $r=2M$, the Kretschmann invariant
is finite so such a singularity can be removed (by an adequate coordinate
transformation). However, $r=0$ is a physical, or not removable, singularity
due to {the fact the} Kretschmann invariant is divergent. Therefore, we point out the
nature of the singularities $r=0$ and $r=2M$ (event horizon) remains
unchanged.}

\section{Some Classical Tests\label{CT}}

{{}In this section, we shall study the motion of particles
in a spacetime described by the spherically symmetric solution (\ref{line}).
With the aim to impose some upper-bounds for the Lorentz-violating
coefficient $\ell$, we consider the Solar system to study the effects of LV
on precession of perihelia of inner planets, the bending of light around the
Sun and the Shapiro time-delay effect.}

{We consider the motion of test particles along the
geodesics described by $x^{\mu }(\lambda )$ obeying the equation,
\begin{equation}
\frac{d^{2}x^{\mu }}{d\lambda ^{2}}+\Gamma _{\,\,\sigma \nu }^{\mu }\frac{%
dx^{\sigma }}{d\lambda }\frac{dx^{\nu }}{d\lambda }=0\text{,}  \label{gdsc}
\end{equation}%
where $\lambda $ is an affine parameter. However, due to the metric
compatibility, it is always possible to use a constant of motion, $\chi $%
, defined by
\begin{equation}
\chi =-g_{\mu \nu }U^{\mu }U^{\nu }\text{,}  \label{const}
\end{equation}%
where the vector $u^{\mu }$ is defined as
\begin{equation}
U^{\mu }=\frac{dx^{\mu }}{d\lambda }\equiv \dot{x}^{\mu }\text{,}
\end{equation}%
where dot denotes differentiation with respect to the affine
parameter. For massive particles, the affine parameter is typically chosen
to be the proper time $\tau $ and $\chi =+1$ (timelike geodesics). On the
other hand, for massless particles we have $\chi =0$ and the parameter $%
\lambda $ is not fixed (null geodesics).}

\subsection{Advance of the perihelion}

{{}From the geodesic equation (\ref{gdsc}), we obtain the
equations describing the trajectory of the massive test particle moving in
the spacetime (\ref{line}):}
\begin{eqnarray}
\frac{d}{d\tau }\left[ \left( 1-\frac{2M}{r}\right) \dot{t}\right] &=&0\text{%
,}  \label{g1} \\[0.08in]
\ddot{r}+\frac{M\left( r-2M\right) }{r^{3}\left( \ell +1\right) }\dot{t}^{2}-%
\frac{M}{r\left( r-2M\right) }\dot{r}^{2} &&  \notag \\
-\frac{r-2M}{\ell +1}\left( \dot{\theta}^{2}+\sin ^{2}\theta \dot{\phi}%
^{2}\right) &=&0\text{,}  \label{g2} \\[0.08in]
\frac{d}{d\tau }\left( r^{2}\dot{\theta}\right) -r^{2}\sin \theta \cos
\theta \dot{\phi}^{2} &=&0\text{,}  \label{g3} \\[0.08in]
\frac{d}{d\tau }\left( r^{2}\sin ^{2}\theta \dot{\phi}\right) &=&0\text{.}
\label{g4}
\end{eqnarray}%
{{}By considering the initial conditions in $\tau =\tau
_{0} $ on the coordinate $\theta $: $\theta \left( \tau _{0}\right) =\frac{%
\pi }{2}$ and $\dot{\theta}\left( \tau _{0}\right) =0$, from Eq. (\ref{g3}),
it follows that $\ddot{\theta}\left( \tau \right) $ and any other higher
order derivatives are equal to zero, so the particle motion is confined to
the plane $\theta =\frac{\pi }{2}$. Therefore, we have a spherically
symmetric spacetime with two Killing vectors corresponding to the conserved
energy (E) and the conserved angular momentum (L). The timelike Killing
vector, $K^{\mu }=\left( \partial _{t}\right) ^{\mu }$, is related to the
conserved particle energy given by}
\begin{equation}
E=-g_{\mu \nu }K^{\mu }U^{\nu }=\left( 1-\frac{2M}{r}\right) \dot{t}\text{,}
\label{E}
\end{equation}%
{{}The rotational Killing vector, $\psi ^{\mu }=\left(
\partial _{\phi}\right) ^{\mu }$, providing the conserved angular momentum
of the particle,}
\begin{equation}
L=g_{\mu \nu }\psi ^{\mu }U^{\nu }=r^{2}\dot{\phi}\text{,}  \label{L}
\end{equation}%
Clearly, the Eqs. (\ref{E}) and (\ref{L}) are consistent with the Eqs. (\ref%
{g1}) and (\ref{g4}), respectively.

{Then, from the conserved quantities in Eq. (\ref{const}%
) for timelike geodesics it yields a single differential equation for the
coordinate $r$ in terms of the proper time $\tau $,
\begin{equation}
\left( 1+\ell \right) \dot{r}^{2}+\left( 1-\frac{2M}{r}\right) \left( \frac{%
L^{2}}{r^{2}}+1\right) =E^{2}\text{.}  \label{gg2}
\end{equation}%
We now introduce the variable $u=r^{-1}$, such that
\begin{equation}
\dot{r}=\frac{dr}{d\phi }\dot{\phi}=-L\frac{du}{d\phi }\text{.}
\end{equation}%
By substituting it in Eq. (\ref{gg2}) we obtain
\begin{equation}
\left( 1+\ell \right) \left( \frac{du}{d\phi }\right) ^{2}+u^{2}=\frac{%
E^{2}-1}{L^{2}}+\frac{2M}{L^{2}}u+2Mu^{3}.  \label{gg3}
\end{equation}%
{As usually done in this treatment}, it is preferable to solve the second-order equation which is obtained by {differentiating the above equation with respect to $\phi$, providing}
\begin{equation}
\left( 1+\ell \right) \frac{d^{2}u}{d\phi ^{2}}+u-\frac{M}{L^{2}}-3Mu^{2}=0%
\text{.}  \label{gg4}
\end{equation}%
It only presents LV contributions into the coefficient of the first term
maintaining the total structure of the one obtained {in the context of} GR. In order to solve
perturbatively the Eq. (\ref{gg4}) and due to the fact we are assuming the LV parameter $%
\ell \ll 1$, it is still valid to consider the last term as a relativistic
correction when compared with the Newtonian case. The perturbative solution
is defined in terms of a small parameter $\epsilon=3M^{2}/L^{2}$:
\begin{equation}
u\simeq u^{\left( 0\right) }+\epsilon u^{\left( 1\right) }.  \label{gg4q}
\end{equation}
}

{The differential equation at zeroth-order in $\epsilon $
yields
\begin{equation}
\left( 1+\ell \right) \frac{d^{2}u^{\left( 0\right) }}{d\phi ^{2}}+u^{\left(
0\right) }-\frac{M}{L^{2}}=0\text{,}
\end{equation}%
whose solution is given by
\begin{equation}
u^{\left( 0\right) }=\frac{M}{L^{2}}\left[ 1+e\cos \left( \frac{\phi }{\sqrt{%
1+\ell }}\right) \right] .  \label{th0}
\end{equation}
It is analogous to the Newtonian result.} Here, the integration constants we
have considered are the orbital eccentricity $e$ (considered small as that
of GR) and the initial value $\phi _{0}=0$.

{The differential equation at first-order in $\epsilon $
is
\begin{equation}
\left( 1+\ell \right) \frac{d^{2}u^{\left( 1\right) }}{d\phi ^{2}}+u^{\left(
1\right) }-\frac{L^{2}}{M}(u^{\left( 0\right) })^{2}=0\text{,}
\end{equation}%
which admits an approximated solution of the form%
\begin{eqnarray}
u^{\left( 1\right) } &\simeq &\frac{M}{L^{2}}e\frac{\phi }{\sqrt{1+\ell }}%
\sin \left( \frac{\phi }{\sqrt{1+\ell }}\right)  \notag \\[0.15cm]
&&+\frac{M}{L^{2}}\left[ \left(1+\frac{e^{2}}{2}\right)-\frac{e^{2}}{6}\cos \left( \frac{%
2\phi }{\sqrt{1+\ell }}\right) \right] \text{.}
\end{eqnarray}%
For our purposes, the second term can be {neglected once it consists of a
constant displacement and a quantity that oscillates around zero.}}

{Therefore, the perturbative solution (\ref{gg4q}) reads
\begin{equation}
u\simeq \frac{M}{L^{2}}\!\left[ 1+e\cos \!\left( \!\frac{\phi }{\sqrt{1+\ell
}}\!\right) +\epsilon e\frac{\phi }{\sqrt{1+\ell }}\sin \!\left( \!\frac{%
\phi }{\sqrt{1+\ell }}\!\right) \right] \text{.}  \label{gg4qq}
\end{equation}%
Because $\epsilon \ll 1$, the perturbative solution (\ref{gg4qq}) can be
rewritten in the form of an ellipse equation,
\begin{equation}
u\simeq \frac{M}{L^{2}}\left[ 1+e\cos \left( \frac{\phi \left( 1-\epsilon
\right) }{\sqrt{1+\ell }}\right) \right] .
\end{equation}%
Despite of the presence of Lorentz violation, the orbit remains periodic
with period $\Phi $,
\begin{equation}
\Phi =\frac{2\pi \sqrt{1+\ell }}{1-\epsilon }\approx 2\pi +\Delta \Phi .
\end{equation}%
The advance of the perihelion ($\Delta \Phi $) is {obtained} by {taking} the
lowest order in the $\epsilon $ and $\ell $ expansion. Then, $\Delta \Phi $ is
given by the following expression
\begin{equation}
\Delta \Phi =2\pi \epsilon +\pi \ell =\Delta \Phi _{\text{GR}}+\delta \Phi _{%
\text{LV}}\text{,}  \label{Ad}
\end{equation}%
where $\Delta \Phi _{\text{GR}}$ is the prediction of GR%
\begin{equation}
\Delta \Phi _{\text{GR}}=2\pi \epsilon =\frac{6\pi G_{N}m}{c^{2}\left(
1-e^{2}\right) a},
\end{equation}%
with $c$ being the speed of light, {$m$} the geometrical mass, {$e$} the orbital eccentricity and $a$ being the semi-major axis of the orbital ellipse .}

{The contribution per period due to the spontaneous
breaking of Lorentz symmetry is given by $\delta \Phi _{\text{LV}}$,
\begin{equation}
\delta \Phi _{\text{LV}}=\pi \ell \text{,}
\end{equation}
The expression (\ref{Ad}) shows the effects of Lorentz violation
identifying it as an additional correction to the GR result.} Thus, since
the perihelion shift of some planetary motions has already been measured
\cite{Shapiro1989,Will1993}, we can establish an estimated attainable
sensitivity for Lorentz violation from the perihelion shifts for the orbit
of the inner planets consistent with the GR.

{Table I shows some values for both the observed and predicted
(by GR) perihelion advance associated to some bodies in the solar system,
according to the most up to date measurement found in Ref. \cite{Pitjeva}.}
Note that, in addition to the planets, we have also presented the
observational data for the asteroid Icarus, which agree with GR predictions
to within $20\%$ of the estimated uncertainty \cite{Weinberg,Shapiro}.

\begin{widetext}

\begin{table}[]{{}
\caption{Theoretical and observed values of perihelion shifts given in
arcseconds per century ($^{\prime \prime }$C$^{-1}$).} \label{TI}\centering%
\begin{tabular}{lccc}
\hline\hline
Planets \ \ \ \ \ \ & \ \ \ \ \ \ \ \ \ \ \ \ \ \ \ \ \ GR prediction$^{a}$ \ \ \ \ \ \ \ \ \ \ \ \ & \ \ \ \ \ \ \ \ \ \ \ \ \ \ \ \ \ \ \ \ \ \  Observed \ \ \ \ \ \ \ \ \ \ \ \ \ \ \ \ \ \ \  \ \ \ \ \ & \ \ \ \ \   Error estimates$^{b}$   \ \ \ \ \   \\ \hline
Mercury (${\mercury}$) & ${42.9814}$ & $\,{42.9794}\,\pm \,0.0030\,$ & $%
-0.0020\,\pm \,0.0030$ \\
Venus (${\venus}$) & \thinspace\ $8.6247$ & \ \ \thinspace\ $\,\,8.6273\,\pm
\,0.0016~~$\ \  & \hspace{0.15cm} $0.0026\,\pm \,0.0016$ \\
Earth (${\oplus }$) & \thinspace\ \thinspace\ $3.83877$ & \ $\,3.83896\,\pm
\,0.00019$ & \hspace{0.15cm} $0.00019\,\pm \,0.00019$ \\
Mars(${\mars}$) & \ \ \ \thinspace\ ${1.350938}$ & \ $\,{1.350918}\,\pm
\,0.000037$ & $-0.000020\,\pm \,0.000037$ \\
Jupiter(${\circledast }$) & \thinspace\ $0.0623$ & \ \thinspace\ $%
\,0.121\,\pm \,0.0283$ & \ \thinspace\ $0.0587\,\pm \,0.0283$ \\
Saturn(${\oslash }$) & \ \ $\ 0.01370$ & \ $\,0.01338\,\pm \,0.00047$ & $%
-0.00032\,\pm \,0.00047$ \\
Icarus & ${10.1}$ \thinspace\ \ \ \ & $\,\ 9.8\,\pm \,0.8$ & ${-0.3}\,\pm \,0.8$ \\ \hline\hline
\end{tabular}
\vspace{-0.15cm}
\par
\begin{flushleft}
{$^{a}$ Computed from the data base of Refs.\cite{planets,Beringer}.}\newline
{$^{b}$ From the Refs.\cite{Pitjeva,Shapiro}.}
\end{flushleft}}
\end{table}

 \end{widetext}

\begin{table}[H]
{{}
\caption{Some upper bounds obtained from the observational error of the
perihelion shifts.} \label{TII}%
\begin{tabular}{cccccccc}
\hline\hline
& Parameters LSB &  &  &  &  & Bounds &  \\ \hline
& $\ell _{{\mercury}}$ &  &  & \hspace{3.5cm} &  & $1.1\times 10^{-11} $ &
\\
& $\ell _{{\venus}}$ &  &  &  &  & $1.5\times 10^{-11} $ &  \\
& $\ell _{{\oplus }}$ &  &  &  &  & $2.9\times 10^{-12} $ &  \\
& $\ell _{{\mars}}$ &  &  &  &  & $1.1\times 10^{-12} $ &  \\
& $\ell _{{\circledast }}$ &  &  &  &  & $5.2\times 10^{-9}$ \, &  \\
& $\ell _{{\oslash }}$ &  &  &  &  & $2.1\times 10^{-10} $ &  \\
& $\ell _{\text{Icarus}}$ &  &  &  &  & $1.3\times 10^{-8}$ \, &  \\
\hline\hline
\end{tabular}%
}
\end{table}

{We use the observational error in experimental data to
compute some upper-bounds for the LV parameter $\ell $. For example, for the
motion of Mercury around the Sun, the observational error is $0.003^{\prime
\prime }$C$^{-1}$ (or $72.3 \times 10^{-7}$ arcseconds per orbit). So, we
suppose the contribution of the Lorentz violation $\delta \Phi _{\text{LV}}$
is less than the observational error. Such a procedure allows us to estimate
an upper-bound at the level of $\ell _{\mercury}< 1.1\times 10^{-11}$. By
applying the same procedure for the other planets in Table \ref{TI}, we have
achieved the set of estimates of attainable experimental sensitivities
(upper-bounds) presented in Table \ref{TII}. Thus, we observe the best
upper-bound attained from the advance of the perihelion is $\ell \lesssim
10^{-12}$.}

\subsection{Bending of light\label{Bending}}

{Unlike the previous case, we now have massless test
particles whose trajectories correspond to null geodesics so $\chi=0$ in Eq.
(\ref{const}) which after substituting the conserved quantities becomes}
\begin{equation}
\left( 1+\ell \right) \dot{r}^{2}+\left( 1-\frac{2M}{r}\right) \frac{L^{2}}{%
r^{2}}=E^{2}\text{,}  \label{Lg}
\end{equation}
{where the dot now denotes differentiation with
respect to some affine parameter.}

Again, we consider $u=r^{-1}$ with $r\equiv r(\phi )$ and the
differentiation with respect to $\phi $ in (\ref{Lg}) which results%
\begin{equation}
\left( 1+\ell \right) \frac{d^{2}u}{d\phi ^{2}}+u-3Mu^{2}=0\text{.}
\label{Lg2}
\end{equation}%
We observe that in the limit $\ell \rightarrow 0$ Eq. (\ref{Lg2}) recovers
the corresponding {{} GR result providing the deflection of
light rays, as expected. In analogy with the previous subsection, we use a
perturbative method to achieve a solution by considering the quantity $Mu$
as sufficiently small. Thus, we write the approximate solution of the form
\begin{equation}
u\simeq u^{\left( 0\right) }+3Mu^{\left( 1\right) },
\end{equation}%
which, {after replacement in Eq. (\ref{Lg2}),} gives the following differential equation
for $u^{(0)}$,
\begin{equation}
\left( 1+\ell \right) \frac{d^{2}u^{\left( 0\right) }}{d\phi ^{2}}+u^{\left(
0\right) }=0\text{,}
\end{equation}%
whose solution is%
\begin{equation}
u^{\left( 0\right) }=\frac{1}{D}\sin \left( \frac{\phi }{\sqrt{1+\ell }}%
\right) \text{,}  \label{Lg3}
\end{equation}%
where $D$ is a constant of integration and we have considered the initial
angle $\phi _{0}=0$, for convenience. {This result corresponds to the equation
of a straight-line which is analogous to the Newtonian prediction.} }

{The differential equation for $u^{(1)}$, {in turn,} becomes
\begin{equation}
\left( 1+\ell \right) \frac{d^{2}u^{\left( 1\right) }}{d\phi ^{2}}+u^{\left(
1\right) }-\frac{1}{D^{2}}\sin ^{2}\left( \frac{\phi }{\sqrt{1+\ell }}%
\right) =0\text{,}  \label{Lg4}
\end{equation}%
and its solution\ is written as%
\begin{equation}
u^{\left( 1\right) }=\frac{1}{3D^{2}}\left[ 1+A\cos \left( \frac{\phi }{%
\sqrt{1+\ell }}\right) +\cos ^{2}\left( \frac{\phi }{\sqrt{1+\ell }}\right) %
\right] .
\end{equation}%
Hence, a general solution for $u(\phi )$ assume the following form%
\begin{eqnarray}
u &\simeq &\frac{1}{D}\sin \left( \frac{\phi }{\sqrt{1+\ell }}\right)
\label{Lg5} \\
&&+\frac{M}{D^{2}}\left[ 1+A\cos \left( \frac{\phi }{\sqrt{1+\ell }}\right)
+\cos ^{2}\left( \frac{\phi }{\sqrt{1+\ell }}\right) \right] \text{,}  \notag
\end{eqnarray}%
with $A$ being an arbitrary constant. }

{Since we are interested in determining the angle of
deflection for a light ray, the boundary conditions are determined by
assuming: $(i)$ The source is located in $r\rightarrow \infty $ such that $%
u(r\rightarrow \infty) \rightarrow 0$ and $\phi =-\delta _{1}$, and $(ii)$
the observer is localized in $r\rightarrow \infty $ such that $%
u(r\rightarrow \infty) \rightarrow 0$ and $\phi =+\delta _{2}$, so the total
angle of deflection is given by $\delta =\delta _{1}+\delta _{2}$. By using
these boundary conditions in Eq. (\ref{Lg5}) and taking in consideration $%
\ell \ll 1$ and $\delta _{1},\delta _{2}\ll 1$, the first-order equation
provides
\begin{eqnarray}
\delta _{1} &=&\frac{M}{D}\left( 2+A\right) , \\[0.15cm]
\delta _{2} &=&\frac{M}{D}\left( 2-A\right) +\frac{\pi \ell }{2}.
\end{eqnarray}%
Hence, the light-ray deflection angle in the metric (\ref{line}) is
\begin{equation}
\delta =\delta _{\text{GR}}+\delta _{\text{LV}}=\frac{4G_{N}m}{c^{2}D}+\frac{%
\pi \ell }{2},  \label{Lg6}
\end{equation}%
with $m$ being the mass of the deflecting body and $D$ {the
so-called} impact parameter (defined as the distance of closest approach of
the light ray to the center of mass of the deflecting body). The first term $%
\delta _{\text{GR}}$,
\begin{equation}
\delta _{\text{GR}}=\frac{4G_{N}m}{c^{2}D},
\end{equation}%
gives the usual deviation of light predicted by the GR. The second term $%
\delta _{\text{LV}}$,%
\begin{equation}
\delta _{\text{LV}}=\frac{\pi \ell }{2},
\end{equation}%
is the correction comming from the LSB effects.} Of course, taking the limit $\ell \rightarrow 0$ in Eq. (\ref{Lg6}) we recover the usual result
established by GR for the bending of light.

For a ray grazing the Sun we have $m=M_{\odot }$ and $D\approx R_{\odot }$ .
Using, for example, the values of Ref. \cite{Beringer}, one can verify that
GR predicts an angle given by $\delta _{\text{GR}}=4G_{N}M_{\odot
}/c^{2}R_{\odot }\approx {{}1.7516687}^{\prime \prime }$. Therefore, if there is,
indeed, Lorentz violation in nature, {{} the effects
arising from the LV term ($\delta _{\text{LV}}$)} must be smaller than the
observational errors. The error bars obtained in recent measurements for the
deflection of light by the Sun \cite{Shapiro2004,LambertI,LambertII,Will2014}
allows us to provide an interesting sensitivity for Lorentz violation.
Specifically, a detailed analysis of the observational data for the bending
of light, adopting the values from Ref. \cite{LambertII} as example, yields
an error bar of order $\sim {{}0.0001051}^{\prime \prime }$. {{}
Taking this value, we set the upper-bound from the inequality: $\delta _{%
\text{LV}}<{{}0.0001051}^{\prime \prime }$. A quickly calculation allows us to found
$\ell <3.2\times 10^{-10}$, it is similar but not better than those
found in the previous test.}

\subsection{Time delay of light}

A further measurable {relativistic phenomenon} involving
light rays is the Shapiro time-delay effect \cite{Shapiro1964}. The
solar-system tests involving this effect can yield interesting sensitivities
to Lorentz violation. {For this purpose we will derive}
an expression involving the Lorentz-violating corrections for time-delay
effect from the result already obtained in the subsection \ref{Bending}.
Namely, we are interested in {an equation providing the
change in the round trip travel time of light to an object due to the}
influence of a massive body such as the Sun.

{By considering the motion of light in the equatorial plane ($\theta =\pi /2$%
) and because it travels along a null geodesic in the spacetime ($\ref{line}$%
), {i.e., the condition $ds^{2}=0$ is satisfied,} we can write}
\begin{equation}
-\left( 1-\frac{2M}{r}\right) dt^{2}+\left( 1+\ell \right) \left( 1-\frac{2M%
}{r}\right) ^{-1}dr^{2}+r^{2}d\phi ^{2}=0\text{.}  \label{TL}
\end{equation}%
{Next, we consider the zero-order solution (\ref{Lg3})
characterizing the straight-line approximation,
\begin{equation}
r\sin \left( \frac{\phi }{\sqrt{1+\ell }}\right) =D,
\end{equation}%
and we use it to establish the following relation,
\begin{equation}
r^{2}d\phi ^{2}=\left( 1+\ell \right) \left( \frac{D^{2}}{r^{2}-D^{2}}%
\right) dr^{2}\text{.}
\end{equation}%
Thus, Eq. (\ref{TL}) can be rewritten as%
\begin{equation}
dt^{2}=\frac{1+\ell }{1-\frac{2M}{r}}\left( \frac{1}{1-\frac{2M}{r}}+\frac{%
D^{2}}{r^{2}-D^{2}}\right) dr^{2}\text{.}
\end{equation}
Expanding it in terms of $M/r$ and considering the contributions at
first-order we get
\begin{equation}
dt\simeq \pm \frac{\sqrt{1+\ell }}{\sqrt{r^{2}-D^{2}}}\left( 1+\frac{2M}{r}-%
\frac{MD^{2}}{r^{3}}\right) rdr.
\end{equation}
}

{The setup for} Shapiro delay effect involves two
stations at large distances from the {{}massive source (or
curvature source)}. {{}By assuming} a light ray (or radar
signal) from an emitter located at $r_{\text{E}}$ traveling to a receiver at
$r_{\text{R}}$, {{}the travel time is given by
\begin{eqnarray}
t &=&t_{0}\sqrt{1+\ell }+2M\sqrt{1+\ell }\ln \left[ \frac{r_{\text{E}%
}+\left( r_{\text{E}}^{2}-D^{2}\right) ^{1/2}}{D}\right]  \notag \\[0.08in]
&&+2M\sqrt{1+\ell }\ln \left[ \frac{r_{\text{R}}+\left( r_{\text{R}%
}^{2}-D^{2}\right) ^{1/2}}{D}\right]  \notag \\[0.08in]
&&-M\sqrt{1+\ell }\left[ \frac{\left( r_{\text{R}}^{2}-D^{2}\right) ^{1/2}}{%
r_{\text{R}}}+\frac{\left( r_{\text{E}}^{2}-D^{2}\right) ^{1/2}}{r_{\text{E}}%
}\right] \!\!,\quad  \label{TL0}
\end{eqnarray}
where $t_{0}$ represents the travel time in flat spacetime,}
\begin{equation}
t_{0}=\left( r_{\text{R}}^{2}-D^{2}\right) ^{1/2}+\left( r_{\text{E}%
}^{2}-D^{2}\right) ^{1/2}\text{,}  \label{TL1}
\end{equation}
It should be noted that in the absence of {{} the Lorentz
violation, $\ell =0$, the Eq. (\ref{TL0})} recovers the expression predicted
by GR, as expected. It is evident that the first term in (\ref{TL0}) stands
the travel of radar signal along a straight-line {{}%
including the effects due to Lorentz violation in a flat spacetime} (special
relativity). The other terms represent {{}the delay
produced by the curved spacetime. Such a delay may be} {{}%
interpreted as an effective increase in the distance between the emitter and
receiver of the radar signal.}

{The result (\ref{TL0}) can be} applied to the
solar-system by considering the reflection of a signal from a planet or
spacecraft. We take, for example, the spacetime near the Sun ($m=M_{\odot }$%
) {{} and a radar} signal emitted from Earth ($r_{\text{E}%
}=r_{\oplus }$) {{} traveling to} a planet or spacecraft {%
{} located at} ($r_{\text{R}}$). For simplicity, {%
{}we consider both the} Earth and the planet as stationary
relative to the Sun. The time-delay is a maximum when the planet is at
superior conjunction {and the} radar signal just {grazes}
the {Sun's} surface such that the radius of closest approach is $D\approx
R_{\odot }$ {{}satisfying the condition $D\ll r_{\oplus
},r_{\text{R}}$. Therefore, from Eq. (\ref{TL0}), the total round-trip time
for a signal traveling from the Earth to an other planet (or spacecraft) and
returning is
\begin{equation}
T\approx T_{0}\sqrt{1+\ell }+\frac{4G_{N}M_{\odot }}{c^{3}}\left[ \ln \left(
\frac{4r_{\oplus }r_{\text{R}}}{R_{\odot }^{2}}\right) -1\right] \sqrt{%
1+\ell }\text{,}  \label{TL2}
\end{equation}%
where $T_{0}$ represents the total travel time in flat spacetime,
\begin{equation}
{T_{0}=\frac{2}{c}\left( r_{\text{R}}^{2}-R_{\odot }^{2}\right) ^{1/2}+\frac{%
2}{c}\left( r_{\oplus }^{2}-R_{\odot }^{2}\right) ^{1/2}}.  \label{TL4}
\end{equation}%
}

{Similarly to {what is done} in GR, from Eq. (\ref{TL2}), we
define, in this Lorentz violating framework, the total excess-delay by
\begin{equation}
\delta T=T-T_{0}=\delta T_{\text{GR}}+\delta T_{\text{LV}}\text{,}
\label{TL3}
\end{equation}%
{where, taking only first order terms in $\ell << 1$, the quantity $\delta T_{\text{GR}}$ is given by}
\begin{equation}
\delta T_{\text{GR}}=\frac{4G_{N}M_{\odot }}{c^{3}}\left[ \ln \left( \frac{%
4r_{\oplus }r_{\text{R}}}{R_{\odot }^{2}}\right) -1\right] \text{,}
\end{equation}%
{and represents} the excess-delay due to pure GR. {The other term,} $\delta T_{\text{LV}}$, is the
contribution of Lorentz violation to the excess-delay,
\begin{equation}
\delta T_{\text{LV}}=\frac{\ell }{2}\left( \delta T_{\text{GR}}+T_{0}\right)\approx\frac{\ell}{2}T_{0},
\end{equation}
and we shall use it to obtain estimates of sensitivities to Lorentz violation.

This could be done from the passive radar measurements of the
inner planets and or active ranging experiments of interplanetary
spacecrafts. For the Shapiro delay with passive radar measurements of the
inner planets, such as Mercury or Venus, used as passive reflectors of the
radar signals, it has been obtained (taking the planet Venus as example) an
excess-delay prediction by general relativity  to well within the experimental
uncertainty of 20\% \cite{Shapiro1968} and, subsequently, within 2\%
\cite{Shapiro1971}.  We take this latter as an upper bound for Lorentz
violating effects which would correspond to a sensitivity of $\ell _{\venus}
<5.0\times 10^{-9}$.

Time delay have also been measured with artificial satellites, such as
Mariner 6 and 7 spacecrafts in orbit around the Sun, used as active
retransmitters of the radar signals. An analysis of the Mariners 6 (M6) and 7 (M7) data suggest that a realistic estimate of the total uncertainty, for both
cases, is perhaps less than 3\% \cite{Anderson1975} so that  the
estimates to  sensitivities for Lorentz violation parameters $\ell _{\text{M6}}$ and $\ell _{\text{M7}}$ are  $2.2\times 10^{-9}$  and $1.6\times 10^{-9}$, respectively .

Another major advance was made using an active transmitter on a spacecraft
stationed on a planet. {{}An example can be given by experiments conducted during the mission of Viking spacecraft to Mars. This consisted of space probes that orbited Mars, equipped with a lander to study the planet from its surface. The measurement from the Viking Mars (VM) landers
resulted} in an estimated accuracy of 0.1\% \cite{Reasenberg1979} which
allows us to establish a sensibility of $\ell _{\text{VM}}< 1.8\times10^{-10}$.

The most precise measurement of the Shapiro time-delay from spacecraft
measurements so far has been made from the Cassini mission during its cruise
to Saturn \cite{Bertotti2003}. Performing a detailed analysis of the data
obtained in the 2002 superior conjunction of Cassini, it is verified the
resulting measurement error must be within at most 0.0012\% of unity. From
this value, we obtain an attainable sensitivity of $\ell _{\text{Cassini}} <
 6.2\times 10^{-13}$.

\section{Conclusions and remarks\label{end}}

{{}We have investigated a static and spherically symmetric vacuum solution which is obtained in the context of a Lorentz-violating modified gravity contained into the framework of a Riemannian bumblebee gravity model.
We have found a new spherically symmetric solution which is very similar to the Schwarzschild one, however,  its Kretschmann invariant (\ref{inv}) guarantees they are very different.

The implications of the theoretical results obtained are studied for some existing classical gravitational experiments, including the advance of the perihelion, bending of light and Shapiro's time-delay.  These tests present an interesting feature: even in the absence of a massive source of curvature we still have corrections coming purely from the Lorentz violation.  Indeed, this is compatible with the Kretschmann invariant (\ref{inv}), which is nonvanishing in the limit $M\rightarrow 0$. This result could indicate the background carrying the Lorentz violating effects also deforms slightly the spacetime, which actually should be approximately flat because of those are significantly small. The smallness of LV effects have allowed to compute some upper-bounds on the parameter $\ell $,  which are all summarized in Table \ref{a}.

In the case of the relativistic perihelion advance, the induced effect by
Lorentz violation can be interpreted as a correction to usual
result of GR, which may be recovered in the limit $\ell\rightarrow0$.
{With the corresponding Lorentz-violating terms at hand, we could estimate attainable sensitivities for some inner planets of the solar system. In this particular scenario, the most stringent upper bound provides $\ell<10^{-12}$.

\begin{table}[H]
\caption{Summary of estimates for upper bounds of radial bumblebee
background in some classical tests.} \label{a}\centering%
\begin{tabular}{lccc}
\hline\hline
{%
\begin{tabular}{@{}c}
Parameters \\[-0.08cm]
LSB%
\end{tabular}%
} \hspace{0.35cm} & \hspace{0.35cm} {%
\begin{tabular}{@{}c}
Advance \\[-0.08cm]
perihelion%
\end{tabular}%
} \hspace{0.35cm} & \hspace{0.35cm} {%
\begin{tabular}{@{}c}
Bending \\[-0.08cm]
light%
\end{tabular}%
} \hspace{0.5cm} & {%
\begin{tabular}{@{}c}
Time \\[-0.08cm]
delay%
\end{tabular}%
} \\ \hline
$\ell _{{\mercury}}$ & $10^{-11}$ & $\cdot \cdot \cdot $ & $\cdot \cdot
\cdot $ \\
$\ell _{{\venus}}$ & $10^{-11}$ & $\cdot \cdot \cdot $ & $10^{-9} $
\\
$\ell _{{\oplus }}$ & $10^{-12}$ & $\cdot \cdot \cdot $ & $%
\cdot \cdot \cdot $ \\
$\ell _{{\mars}}$ & $10^{-12}$ & $\cdot \cdot \cdot $ & $%
\cdot \cdot \cdot $ \\
$\ell _{{\circledast }}$ & $10^{-9}$ \hspace{0.01cm} & $\cdot \cdot \cdot $ & $\cdot \cdot
\cdot $ \\
$\ell _{{\oslash }}$ & $10^{-10}$ & $\cdot \cdot \cdot $ & $\cdot \cdot
\cdot $ \\
$\ell _{\text{Icarus}}$ & $10^{-8}$ \hspace{0.01cm} & $\cdot \cdot \cdot $ & $\cdot \cdot
\cdot $ \\
$\ell $ & $\cdot \cdot \cdot $ & $10^{-10}$ & $\cdot \cdot \cdot $ \\
$\ell _{\text{M6}}$ & $\cdot \cdot \cdot $ & $\cdot \cdot \cdot $ & $10^{-9}$ \\
$\ell _{\text{M7}}$ & $\cdot \cdot \cdot $ & $\cdot \cdot \cdot $ & $10^{-9}$ \\
$\ell _{\text{VM}}$ & $\cdot \cdot \cdot $ & $\cdot \cdot \cdot $ & \hspace{%
0.013cm} $10^{-10}$ \\
$\ell _{\text{Cassini}}$ & $\cdot \cdot \cdot $ & $\cdot \cdot \cdot $ &
\hspace{0.013cm} ${{}10^{-13}}$ \\ \hline\hline
\end{tabular}
\end{table}

The calculation for the bending of light also provides a
Lorentz violating correction to the GR result. Such a LV term allows to
establish an interesting upper bound for the parameter $\ell$. For this
purpose, the analysis was carried out by using the very long baseline
interferometry (VLBI) data \cite{LambertII} providing an attainable
sensitivity at the level of $\ell < 10^{-10}$.

The relativistic effect involving the Shapiro time-delay
of light also yields, as well as the other cases, Lorentz violating
corrections to GR. It follows that, among the data used for the estimates
sensitivities of Shapiro's time-delay effect that might be attainable, the
spacecraft Cassini has provided the most precise measurement at present,
yielding an upper-bound of $\ell < 10^{-13}$ .

Within the context developed in the manuscript, we are exploring other possible vacuum configurations of bumblebee field producing Lorentz violating solutions.  In addition, there exist the possibility of exploring the effect of bumblebee field on black hole solutions such as the charged and rotating ones. The results of these research will be reported elsewhere.}

\begin{acknowledgments}
We thank CAPES, CNPq and FAPEMA (Brazilian agencies) for partial financial
support.
\end{acknowledgments}


\begin{thebibliography}{99}
\bibitem{PRD39.1989} V. A. Kosteleck\'{y} and S. Samuel, Phys. Rev. D
\textbf{39}, 683 (1989); Phys. Rev. Lett. \textbf{63}, 224 (1989); Phys.
Rev. D \textbf{40}, 1886 (1989).

\bibitem{PRD55.1997} D. Colladay and V. A. Kosteleck\'{y}, Phys. Rev. D
\textbf{55}, 6760 (1997).

\bibitem{PRL87.2001} S. M. Carroll, J. A. Harvey, V. A. Kosteleck\'{y}, C.
D. Lane, T. Okamoto, Phys. Rev. Lett. \textbf{87}, 141601 (2001).

\bibitem{LoopQG} Rodolfo Gambini, Jorge Pullin, Phys. Rev. D \textbf{59},
124021 (1999); John Ellis, N. E. Mavromatos, D. V. Nanopoulos, Gen. Rel.
Grav. \textbf{32}, 127 (2000).

\bibitem{PRD69.2004} V. A. Kosteleck\'{y}, Phys. Rev. D \textbf{69}, 105009
(2004).

\bibitem{PLB511.2001} D. Colladay, A. Kosteleck\'{y}, Phys. Lett. B \textbf{%
\ 511}, 209 (2001); V. A. Kosteleck\'{y} and R. Lehnert, Phys. Rev. D
\textbf{63}, 065008 (2001).

\bibitem{EPJP129.2014} K. Bakke and H. Belich, Eur. Phys. J. Plus \textbf{129%
}: 147 (2014).

\bibitem{JMP40.1999} V. A. Kosteleck\'{y} and C. D. Lane, Journal of
Mathematical Physics \textbf{40}, 6245 (1999).

\bibitem{PRD86.2012} T. J. Yoder and G. S. Adkins, Phys. Rev. D \textbf{86},
116005 (2012).

\bibitem{PRD68.2003.2004} R. Lehnert, Phys. Rev. D \textbf{68}, 085003
(2003); J. Math. Phys. \textbf{45}, 3399 (2004).

\bibitem{JMP.2007} O. G. Kharlanov and V. Ch. Zhukovsky, J. Math. Phys.
\textbf{48}, 092302 (2007).

\bibitem{KM} V. A. Kosteleck\'{y} and M. Mewes, Phys. Rev. Lett. \textbf{87}%
, 251304 (2001); Phys. Rev. D \textbf{66}, 056005 (2002); Phys. Rev. Lett.
\textbf{97}, 140401 (2006); Phys. Rev. Lett. \textbf{87}, 251304 (2001)

\bibitem{CFJ} S.M. Carroll, G.B. Field and R. Jackiw, Phys. Rev. D \textbf{41%
}, 1231 (1990); C. Adam and F. R. Klinkhamer, Nucl. Phys. B \textbf{607},
247 (2001); Nucl. Phys. B \textbf{657}, 214 (2003).

\bibitem{n11} A. Moyotl et al., Int. J. Mod. Phys. A \textbf{29}, 1450107
(2014); Int. J. Mod.Phys. A \textbf{29}, 1450107 (2014).

\bibitem{n12} W. F. Chen and G. Kunstatter, Phys. Rev. D \textbf{62}, 105029
(2000); C. D. Carone, M. Sher, and M. Vanderhaeghen, Phys. Rev. D \textbf{74}%
, 077901 (2006).

\bibitem{n13} F.R. Klinkhamer and M. Schreck, Nucl. Phys. B \textbf{848}, 90
(2011); M. Schreck, Phys. Rev. D \textbf{86}, 065038 (2012); M. A. Hohensee,
R. Lehnert, D. F. Phillips, and R. L. Walsworth, Phys. Rev. D \textbf{80},
036010 (2009).

\bibitem{PLB628.2005} B. Altschul and V. A. Kosteleck\'{y}, Phys. Lett. B
\textbf{628}, 106 (2005).

{{}
\bibitem{EWeak} D. Colladay and P. McDonald, Phys. Rev. D \textbf{79}, 125019
(2009); V. E. Mouchrek-Santos and M. M. Ferreira, Jr., Phys. Rev. D \textbf{95}, 071701(R) (2017)

\bibitem{Strong} D0 Collaboration (V. M. Abazov (Dubna, JINR) et al.), Phys. Rev. Lett. \textbf{108}, 261603 (2012)

\bibitem{Hadron} M. S. Berger, V. A. Kosteleck\'{y}, and Z. Liu, Phys. Rev. D \textbf{93}, 036005 (2005) }

\bibitem{n14} R. Bluhm, V. Alan Kosteleck\'{y}, Phys. Rev. D \textbf{71},
065008 (2005).

\bibitem{n16} A. F. Santos et al, Mod. Phys. Lett. A \textbf{30}, 1550011
(2015).

\bibitem{n17} R. V. Maluf, Victor Santos, W. T. Cruz, and C. A. S. Almeida
Phys. Rev. D \textbf{88}, 025005 (2013).

\bibitem{n18} Maluf, R.V. et al. Phys. Rev. D \textbf{90} (2014) no.2,
025007.

\bibitem{PRD74.2006} Q. G. Bailey and V. A. Kosteleck\'{y}, Phys. Rev. D
\textbf{74}, 045001 (2006).

\bibitem{PRD77.2008} R. Bluhm, S.-H. Fung, and V. A. Kosteleck\'{y}, Phys.
Rev. D \textbf{77}, 065020 (2008).

{{}
\bibitem{Gwaves} V. A. Kosteleck\'{y}, A. C. Melissinos and M. Mewes, Phys. Lett. B \textbf{761}, 1-7 (2016); V. A. Kosteleck\'{y} and M. Mewes, Phys. Lett. B \textbf{757}, 510-514 (2016).}

\bibitem{PRD71.2005} R. Bluhm and V. A. Kosteleck\'{y}, Phys. Rev. D \textbf{%
71}, 065008 (2005).

\bibitem{PRD79.2009} V. A. Kosteleck\'{y} and R. Potting, Phys. Rev. D
\textbf{79}, 065018 (2009).

\bibitem{PRD90.2014} R. V. Maluf, C. A. S. Almeida, R. Casana and M. M.
Ferreira, Jr., Phys. Rev. D \textbf{90}, 025007 (2014).

\bibitem{GRG37.2005} V. A. Kosteleck\'{y} and R. Potting, Gen. Relativ.
Gravit. \textbf{37}, 1675 (2005).

\bibitem{PRD72.2005} O. Bertolami and J. P\'{a}ramos, Phys. Rev. D \textbf{72%
}, 044001 (2005).

\bibitem{Carroll} S. M. Carroll, \emph{Spacetime and Geometry: An
Introduction to General Relativity} (Addison-Wesley, San Franscisco, 2004).

\bibitem{Weinberg} S. Weinberg, \emph{Gravitation and Cosmoly: Principles
and Applications of the General Theory of Relativity} (John Wiley \& Sons,
New York, 1972).

\bibitem{planets} The recommended orbital value could be found at \href{https://nssdc.gsfc.nasa.gov/planetary/factsheet}%
{Planetary Fact Sheet - NASA}, accessed 29 May 2017; For the asteroid Icarus
we have used the data of \href{https://ssd.jpl.nasa.gov/sbdb.cgi?orb=1;sstr=1566}%
{JPL Small-Body Database - NASA}, accessed 01 Jun 2017.

\bibitem{Beringer} J. Beringer et al. (Particle Data Group), Phys. Rev. D
\textbf{86}, 010001 (2012).

\bibitem{Shapiro1989} See, for example, I. I. Shapiro, "Solar system tests
of GR: Recent results and present plans", \emph{General Relativity and
Gravitation: Proc. 12th Int. Conf. General Relativity and Gravitation,
University of Colorado at Boulder}, July 2-8, 1989, eds. N. Ashby, D. F.
Bartlett and W. Wyss (Cambridge University Press, Cambridge, 1990), pp.
313-330.

\bibitem{Will1993} For an interesting review, see C. M. Will, Theory and
Experiment in Gravitational Physics (Cambridge University Press, Cambridge,
England, 1993).

\bibitem{Pitjeva} N. P. Pitjev and E. V. Pitjeva, Astron. Lett. \textbf{39},
141 (2013); Mon. Not. R. Astron. Soc. \textbf{432}, 3431 (2013).

\bibitem{Shapiro} I. I. Shapiro, M. E. Ash, and W. B. Smith, Phys. Rev.
Lett. \textbf{20}, 1517 (1968); I. I. Shapiro, W. B. Smith, M. E. Ash, and
S. Herrick, Astron. J. \textbf{76}, 588 (1971).

\bibitem{Shapiro2004} S. S. Shapiro, J. L. Davis, D. E. Lebach and J. S.
Gregory, Phys. Rev. Lett., \textbf{92}, 121101 (2004).

\bibitem{LambertI} S. B. Lambert and C. Le Poncin-Lafitte, Astron.
Astrophys., \textbf{499}, 331-335 (2009).

\bibitem{LambertII} S. B. Lambert and C. Le Poncin-Lafitte, Astron.
Astrophys., \textbf{529}, A70 (2011).

\bibitem{Will2014} See for example, C. M. Will, Living Rev. Relativ. \textbf{%
17}, 4 (2014); updated 1403.7377 [gr-qc].

\bibitem{Shapiro1964} I. I. Shapiro, Phys. Rev. Lett. \textbf{13}, 789
(1964).

\bibitem{Shapiro1968} I. I. Shapiro et al., Phys. Rev. Lett. \textbf{20},
1265 (1968);

\bibitem{Shapiro1971} I. I. Shapiro et al., Phys. Rev. Lett. \textbf{26},
1132 (1971);

\bibitem{Anderson1975} J. D. Anderson, P. B. Esposito, W. Martin, \DH
\textexclamdown\ L. Thornton, and D. O. Muhleman, Ap. J. \textbf{200}, 221
(1975).

\bibitem{Reasenberg1979} R. D. Reasenberg et al., Ap. J. \textbf{234}, 219
(1979).

\bibitem{Bertotti2003} B. Bertotti, L. Iess and P. Tortora, Nature, \textbf{%
425}, 374 (2003).
\end{thebibliography}
\end{document}